\documentclass[aps,pra,reprint,superscriptaddress,showpacs,showkeys,floatfix]{revtex4-1}

\usepackage{graphicx}
\usepackage{color}
\usepackage{amssymb}
\usepackage{amsmath}
\usepackage{epsfig}
\usepackage{bm}
\usepackage[american]{babel}
\usepackage{hyperref}
\usepackage{braket}

\usepackage[T1]{fontenc} 
\usepackage{txfonts}

\DeclareGraphicsExtensions{.pdf,.png,.jpg}

\hypersetup{colorlinks=true,urlcolor=blue,breaklinks=true,citecolor=blue,linkcolor=blue,pdfstartview=FitH,pdfpagemode=UseNone}

\newcommand{\note}[1]{{\color{black}{#1}}}

\newcommand{\intmeasure}{\mathrm{d}^3 r}   

\newcommand{\ev}[1]{\langle #1 \rangle}         
\newcommand{\ve}[1]{\mathbf{#1}}                 
\newcommand{\gF}{g_{F}}
\newcommand{\muB}{\mu_{\textrm{B}}}
\newcommand{\xin}{\xi^{\textrm{C}}_{\textrm{n}}}
\newcommand{\xis}{\xi^{\textrm{C}}_{\textrm{s}}}
\newcommand{\xih}{\xi^{\textrm{BN}}_{\textrm{h}}}
\newcommand{\xiv}{\xi^{\textrm{BN}}_{\textrm{v}}}
\newcommand{\aB}{a_{\textrm{B}}}

\begin{document}

\title{Three-dimensional skyrmions in spin-2 Bose--Einstein condensates}
\date{\today}

\author{Konstantin Tiurev}  \email{konstantin.tiurev@gmail.com}
\affiliation{QCD Labs, QTF Centre of Excellence, Department of Applied Physics, Aalto University, P.O. Box 13500, FI-00076 Aalto, Finland}
\author{Tuomas Ollikainen}
\affiliation{QCD Labs, QTF Centre of Excellence, Department of Applied Physics, Aalto University, P.O. Box 13500, FI-00076 Aalto, Finland}
\author{Pekko Kuopanportti}
\affiliation{Department of Physics, University of Helsinki, P.O. Box 43, FI-00014 Helsinki, Finland}
\affiliation{School of Physics and Astronomy, Monash University, Victoria 3800, Australia}
\author{Mikio Nakahara} 
\affiliation{Department of Mathematics, Shanghai University, 99 Shangda Road, Shanghai 200444, P. R. China}
\affiliation{QCD Labs, QTF Centre of Excellence, Department of Applied Physics, Aalto University, P.O. Box 13500, FI-00076 Aalto, Finland}
\affiliation{Department of Physics, Kindai University, Higashi-Osaka 577-8502, Japan}
\author{David S. Hall} 
\affiliation{Department of Physics and Astronomy, Amherst College, Amherst, Massachusetts 01002-5000, USA}
\author{Mikko M{\"o}tt{\"o}nen} 
\affiliation{QCD Labs, QTF Centre of Excellence, Department of Applied Physics, Aalto University, P.O. Box 13500, FI-00076 Aalto, Finland}
\affiliation{University of Jyv{\"a}skyl{\"a}, Department of Mathematical Information Technology, P.O. Box 35, FI-40014 University of Jyv{\"a}skyl{\"a}, Finland}

\begin{abstract}
We introduce topologically stable three-dimensional skyrmions in the cyclic and biaxial nematic phases of a spin-2 Bose--Einstein condensate. These skyrmions exhibit exceptionally high mapping degrees resulting from the versatile symmetries of the corresponding order parameters. We show how these structures can be created in existing experimental setups and study their temporal evolution and lifetime by numerically solving the three-dimensional Gross--Pitaevskii equations for realistic parameter values. \note{Although the biaxial nematic and cyclic phases are observed to be unstable against transition towards the ferromagnetic phase, their lifetimes are long enough for the skyrmions to be imprinted and detected experimentally.} 
\end{abstract}

\maketitle

\section{\label{sc:intro}Introduction}

Topological defects in spinor Bose--Einstein condensates (BECs) have been the subject of many theoretical and experimental studies over the past decade~\cite{Stamper-Kurn2013,Kaw2012.PhysRep520.253,Ued2014.RepProgPhys77.122401,RevModPhys.81.647}. 
The types of supported defects are determined by the topological properties of the order parameter space in a given phase and classified by its homotopy groups~\cite{Mer1979.RMP51.591,Nak2003_book}. While BECs in the spin-1 state only exhibit polar and ferromagnetic ground-state phases, spin-2 condensates permit a rich variety of phases and, consequently, offer a platform for various new kinds of topological objects~\cite{Ueda.2002.Phys.Rev.A.65.063602,Makela.2003.JPA.36.32,Ued2014.RepProgPhys77.122401}. 

The experimental detection of quantised vortices has become routine in studies of superfluidity in gaseous BECs: the types of observed line defects in three dimensions include singly and multiply quantised vortices~\cite{Lovegrove.PRA.93.033633,Mat1999.PRL83.2498,Mad2000.PRL84.806,Lea2002.PRL89.190403,Iso2007.PRL99.200403}, coreless~\cite{Lea2003.PRL90.140403,Les2009.PRL103.250401}, polar-core~\cite{Sad2006.Nat443.312}, solitonic~\cite{Don2014.PRL113.065302} and half-quantum vortices~\cite{Seo.Phys.Rev.Lett.115015301}, and vortex lattices~\cite{Abo2001.Sci292.476,Ji2008.PRL101.010402}. The interest in topological defects of dimensionality other than one has grown in recent years: the creation of two-dimensional skyrmions~\cite{Cho2012.PRL108.035301,Les2009.PRL103.250401,Cho2012.NJP14.053013} was followed by the observation of point-like defects analogous to Dirac~\cite{Ray2014.Nat505.657,PhysRevX.7.021023} and 't~Hooft--Polyakov~\cite{Ray2015.Sci348.544} monopoles. 

Three-dimensional skyrmions and knots are \note{topological field configurations} classified by the nontrivial elements of the third homotopy group $\pi_3$. 
Knots are identified by an $S^3 \rightarrow S^2$ mapping and characterized by a linking number. In particular, a knot soliton consists of an infinite number of linked loops, each corresponding to a distinct point of the order parameter space\note{~\cite{Ued2014.RepProgPhys77.122401,Kaw2008.PRL100.180403, PhysRevLett.81.4798}}.
The first experimental observation of knot solitons with unit linking number was achieved by Hall \emph{et al.}~\cite{Hal2016.NatPhys12.478} by imprinting the topological structure into the order parameter of a polar-phase spin-1 BEC. Three-dimensional skyrmions, on the other hand, are identified by an $S^3\rightarrow G$ mapping and characterized by a mapping degree that counts the number of times the order parameter space $G$ is covered. Originally introduced by Skyrme in a classical field theory~\cite{Skyrme.Proc.R.Soc.A260.127.1961}, the skyrmionic textures are predicted to appear in condensed matter systems such as superfluid $^3$He-A~\cite{JETP.46.2.1977,Shankar.J.Phys.France.38.1405,And1977.PRL38.508}, liquid crystals~\cite{RevModPhys.61.385}, quantum Hall systems~\cite{PhysRevLett.75.4290}, and \note{multicomponent BECs~\cite{PhysRevLett.86.3934,Savage.Phys.Rev.Lett.91010403,Liu.Phys.Lett.A.377,PhysRevA.64.043612,Kaw2012.PhysRep520.253}}. Despite their long history and ubiquity, three-dimensional skyrmions have  been observed experimentally only very recently~\cite{Leeeaao3820}, in the form of Shankar skyrmions~\cite{Shankar.J.Phys.France.38.1405} in a ferromagnetic spin-1 BEC.

Spin-2 BECs exhibit two magnetic phases that have no spin-1 counterparts, namely, the \emph{biaxial nematic} (BN) and the \emph{cyclic} (C) phases.
To date, research on topological defects in these phases has been restricted to the study of surface solitons~\cite{Les2009.PRL103.250401} and non-Abelian vortices, the latter of which exhibit noncommutative reconnection dynamics~\cite{Kob2009.PRL103.115301,Kobayashi.J.Phys.Conf.Ser.297.012013,Borgh.Phys.Rev.Lett.117275302} predicted to result in exotic quantum turbulence~\cite{Mawson.Phys.Rev.A91063630}. \note{However, there have been no detailed studies of three-dimensional skyrmions in these phases. The existence of such skyrmions is permitted by the nontrivial nature of the third homotopy groups of both the BN and the C order parameter spaces~\cite{KOBAYASHI2012577,Kaw2012.PhysRep520.253}. In this paper, we introduce exotic skyrmion configurations that are found to exhibit high mapping degrees $Q_{\textrm{BN}}=16$ and $Q_{\textrm{C}}=24$ for the BN and C phases, respectively.} We simulate their dynamical creation, discuss their topological properties, and examine their lifetimes, which are ultimately limited by the stability of the underlying magnetic phase. Our numerical results suggest that it is possible to create and observe these unique topological entities with currently available experimental setups.

The remainder of this article is organised as follows. In Sec.~\ref{sc:theory}, we characterize the symmetries of the mean-field ground states by employing the Majorana representation and formulate the Gross--Pitaevskii equation that governs the temporal evolution of the spin-2 BEC. Section~\ref{sc:methods} describes the method for creating the skyrmions, which are subsequently analyzed in Sec.~\ref{sc:results}. Finally, we provide our conclusions in Sec.~\ref{sc:summary}.

\section{\label{sc:theory}Magnetic phases in a spin-2 condensate}
\subsection{Mean-field ground states of a spin-2 condensate}

A spin-$F$ atomic BEC can be described by a mean-field order parameter that takes a vectorial form with $2F+1$ components. Specifically, we write the order parameter field of a spin-2 condensate as $\Psi(\mathbf{r}) = e^{i\varphi(\mathbf{r})}\sqrt{n(\mathbf{r})} \mathbf{\xi}(\mathbf{r})$, where $n = \Psi^{\dagger}\Psi$ is the particle density, $\varphi$ is the global phase, and $\mathbf{\xi} \in \mathbb{C}^5$ is a normalized spinor obeying $\mathbf{\xi}^{\dagger}\mathbf{\xi} = 1$. In order to understand the different types of ground-state magnetic phases of a spin-2 BEC in the absence of an external magnetic field, it is sufficient to consider the minimization of the spin-dependent interaction energy functional
\begin{eqnarray}
E_{\textrm{int}}[\Psi] = \int \frac{n^2(\mathbf{r})}{2} \Big{\{}c_1|\mathbf{S}(\mathbf{r})|^2 + c_2|A_{20}(\mathbf{r})|^2 \Big{\}}\mathrm{d}^3 r,
\label{eq:energy}
\end{eqnarray}
where $c_1$ and $c_2$ are material constants and $\mathbf{S} = \xi^{\dagger}\mathbf{F}\xi$ is the spin vector based on spin-2 matrices satisfying $[F_a,F_b] = i\epsilon_{abc}F_c$; here $\epsilon_{abc}$ is the Levi-Civita symbol and $\textrm{\textit{a},\textit{b},\textit{c}} \in \{x,y,z\}$. In comparison to the spin-1 case, the interaction energy includes an additional term $c_2|A_{20}|^2/2$, where 
\begin{align}
A_{20} = \frac{1}{\sqrt{5}}(2\xi_2\xi_{-2} - 2\xi_1\xi_{-1}+\xi_0^2) \equiv a_{20} / \sqrt{5} 
\end{align}
is the amplitude of the spin-singlet pair. We are particularly interested in phases absent in spin-1 condensates: the C phase characterized by $(|\mathbf{S}| = 0, |a_{20}| = 0)$ and the BN phase characterized by $(|\mathbf{S}| = 0, |a_{20}| = 1)$.

The BN phase, which minimizes $E_\mathrm{int}[\Psi]$ for $c_1>0$ and $c_2<0$, is represented by the reference spinor
\begin{align} 
\xih = (1,0,0,0,1)^{\textrm{T}}/\sqrt{2}. \label{eq:BN-h}
\end{align}
In the absence of an external magnetic field, the BN phase and the uniaxial nematic (UN) phase, $\xi^{\textrm{UN}} = (0,0,1,0,0)^{\textrm{T}}$, are energetically degenerate within the mean-field theory~\cite{PhysRevLett.98.190404,PhysRevLett.98.160408}, and can be distinguished by the amplitude of spin-singlet trio formation, $|A_{30}|^2 = |3\sqrt{6}(\zeta_2 \zeta_{-1}^2 + \zeta_1^2 \zeta_{-2}) + \xi_0(\xi_0^2 -3\xi_1\xi_{-1} - 6\xi_2\xi_{-2})/2|^2$~\cite{Kaw2012.PhysRep520.253,PhysRevLett.105.230406}.

The C phase appears as the ground state when both $c_1$ and $c_2$ are positive. In this case, the representative spinor minimizing the energy functional $E_{\mathrm{int}}[\Psi]$ can be written as 
\begin{align}
\xin = (\sqrt{1/3},0,0,\sqrt{2/3},0)^{\textrm{T}}. \label{eq:C-n}
\end{align}
This phase minimizes the spin length and simultaneously breaks the spin-related time-reversal symmetry, i.e., ${\xin}^{\dagger} \mathcal{T} \xin = 0$. Here, $\mathcal{T}$ is the time-reversal operator such that~\cite{Kaw2012.PhysRep520.253}
\begin{equation}\label{eq:timerev}
\mathcal{T}\Psi_{m_{\textrm{z}}} = (-1)^{m_{\textrm{z}}}\Psi^*_{-m_{\textrm{z}}},
\end{equation}
where ${m_{\textrm{z}}}\in\left\{2,1,\dots,-2\right\}$ is the spin index in the $z$-quantised basis. Thus the C phase has no analog in the spin-1 case.

\begin{figure}[t]	
\includegraphics[width=0.75\columnwidth]{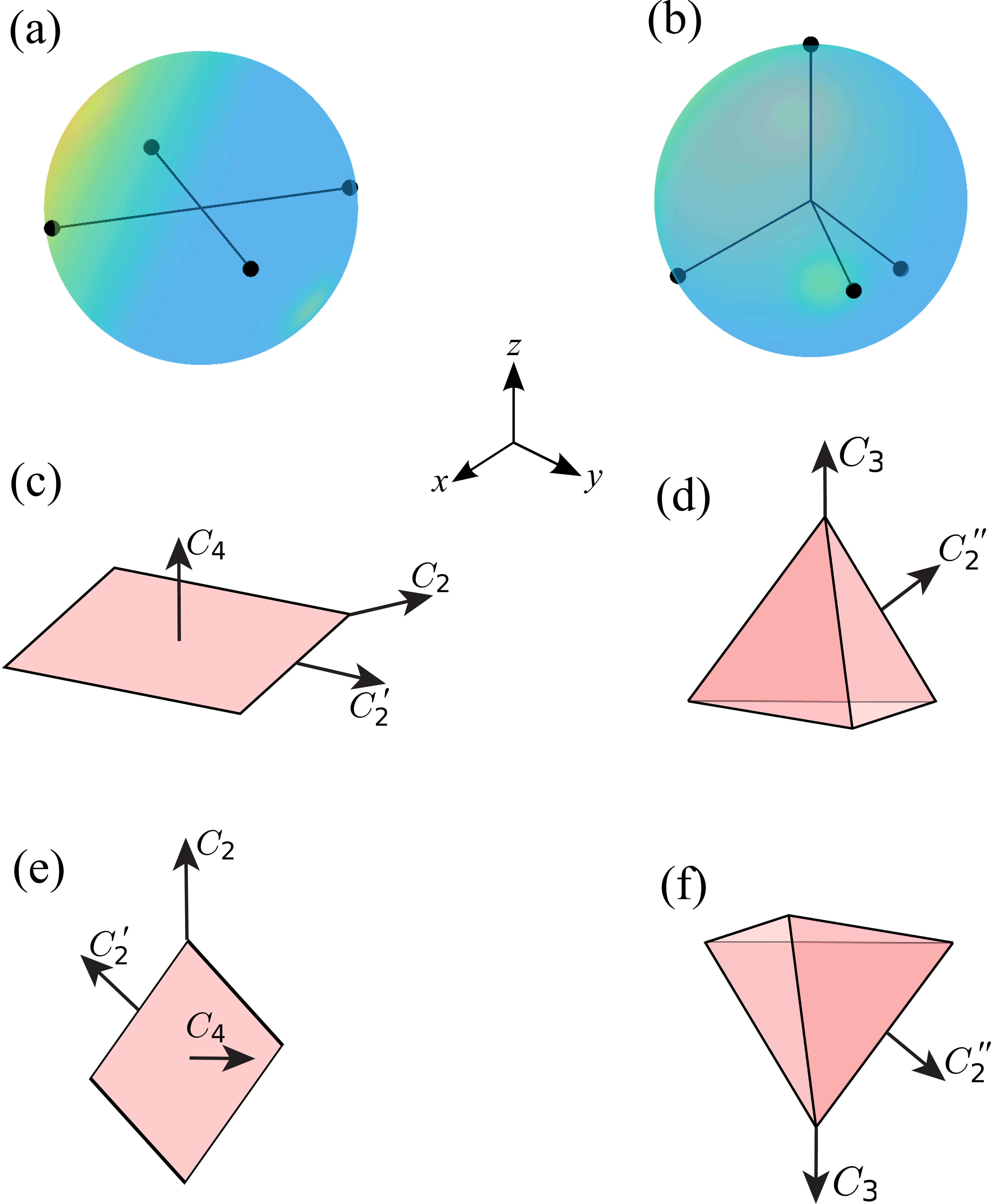}
\caption{\label{fig:1} (a)~Biaxial nematic and (b)~cyclic order parameters represented as symmetric configurations of spin-$1/2$ states shown with black dots ($\bullet$) on the Bloch sphere. Panels (c)--(f) explicitly show the (c)~horizontal and (e)~vertical orientations of the BN state, (d)~north-pole and (f)~south-pole orientations of the C state, and the corresponding symmetry axes. Here, the vertices correspond to the black dots in (a) and (b). The orientation of the coordinate axes is the same for all panels.}
\end{figure}

\subsection{Majorana representation}

In order to characterize the symmetry of the order parameter, it is convenient to employ a geometric representation, sometimes also referred to as the Majorana representation~\cite{Majorana1932,Zho2001.PRL87.080401,PhysRevLett.97.180412,PhysRevA.84.053616}. Here, we express the spin-2 state as totally symmetrised four spin-$1/2$ states, which we associate with Bloch vectors on the 2-sphere. The polar coordinates $(\theta, \phi)$ of the Bloch vectors are given by the stereographic mapping $u=\tan(\theta/2)e^{i\phi}$, where $u$ is one of the four roots of the polynomial 
\begin{equation}
\begin{aligned}
P[\xi](u) &= \sum_{k=0}^{4}\xi^{*}_{k-2}
\sqrt{
\begin{pmatrix}
4\\
k
\end{pmatrix}}
u^k
\\& = \xi_2^*u^4 + 2\xi_1^*u ^3 + \sqrt{6}\xi_0^*u^2 + 2\xi_{-1}^*u + \xi_{-2}^{*}.
\end{aligned}
\end{equation}


The Majorana representation of the BN spinor $\xih$ of Eq.~\eqref{eq:BN-h} is shown schematically in Fig.~\ref{fig:1}(a). This state breaks the cylindrical symmetry of the polar phase into the discrete symmetry of a square, yielding the order parameter space $G_{\textrm{BN}} = \textrm{U}(1)\times\textrm{SO}(3)/D_4$, where $D_4$ is the dihedral group of order 4. In particular, each of the orientations shown in Figs.~\ref{fig:1}(c) and \ref{fig:1}(e) corresponds to all Bloch vectors lying in the $xy$ plane. Owing to their geometric orientation, we will refer to such states as \emph{horizontal} states. We further define another configuration of the BN spinor, obtained from $\xih$ as
\begin{equation}
\begin{aligned}
\xiv &= -i \exp\big{(}-i\frac{\pi}{2}F_y\big{)}
\exp\big{(}-i\frac{\pi}{4}F_z\big{)} \xih 
\\& = (0,1,0,1,0)^{\textrm{T}}/\sqrt{2},
\label{eq:BN-v}
\end{aligned}
\end{equation}
which is the representative spinor of the \emph{vertical} state corresponding to two of the Bloch vectors being aligned with the $z$ axis as in Fig.~\ref{fig:1}(e).

\begin{figure}[t]	
\includegraphics[width=0.85\columnwidth]{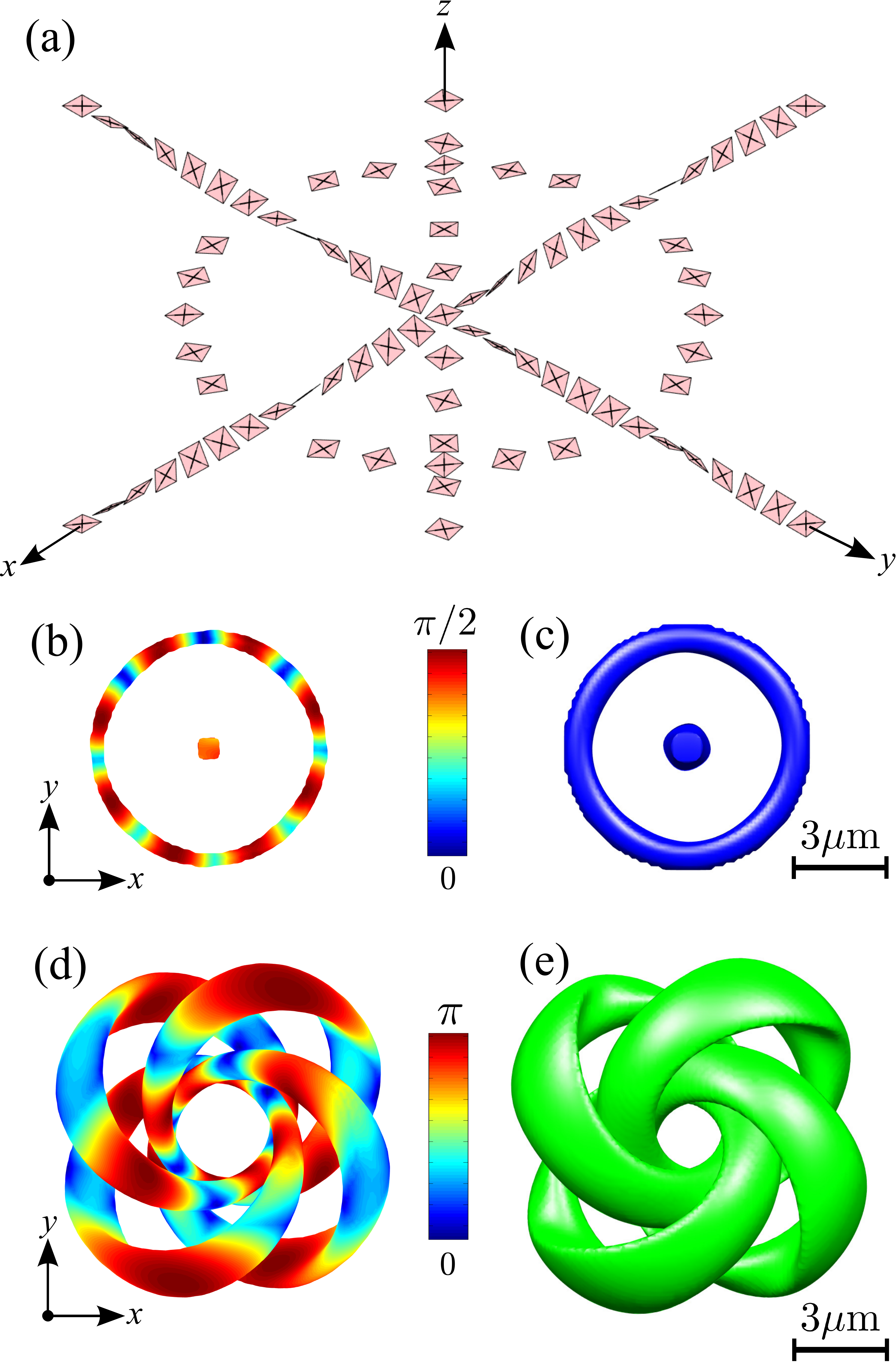}
\caption{\label{fig:2} Skyrmion-like state observed after $0.5$~ms of rotation of the horizontal biaxial nematic spinor $\xih$. Panel (a) schematically shows the orientations of the order parameter along the three coordinate axes and the ring formed by the $\pi$ rotation of the initial state in the $xy$ plane. Preimages of horizontal and vertical states are shown in (b) and (d), respectively. The preimage of the horizontal state in (b) consists of the inner ring in the plane $z=0$ (shown) and all the points corresponding to $2\pi$ rotations that constitute the boundary of the skyrmion (not shown). The color represents the rotation angle about the symmetry axes (b) $C_4$ and (d) $C_2$. Panels (c) and (e) show the isosurfaces $|\xi_2|^2 = 0.4$ and $|\xi_{-1}|^2 = 0.4$, respectively. All images correspond to $t=0.5$~ms after the start of the nonadiabatic creation ramp described in Sec.~\ref{sc:methods}.} 
\end{figure}   
\begin{figure}[t]	
\includegraphics[width=0.85\columnwidth]{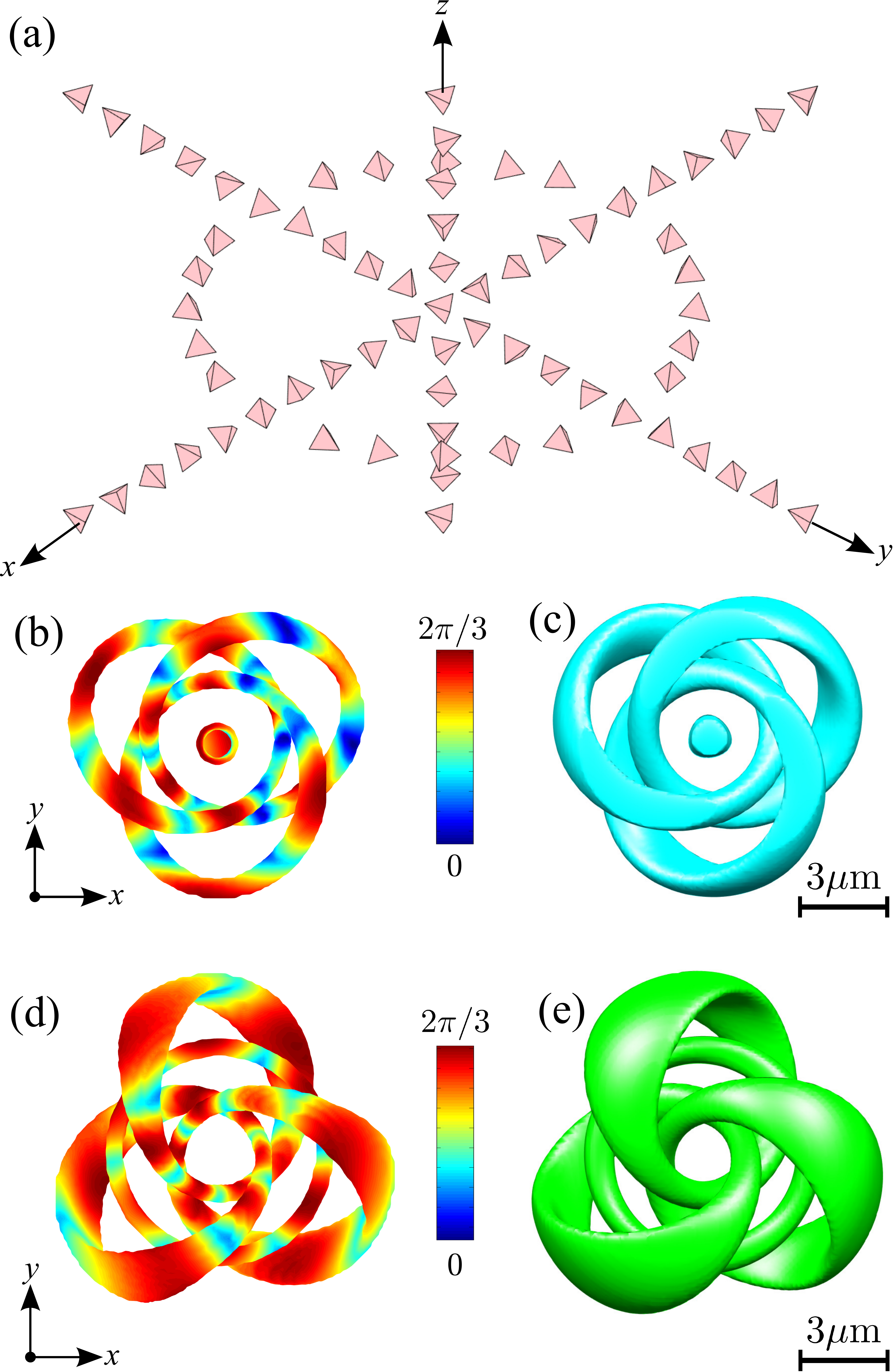}
\caption{\label{fig:3} Skyrmion-like state observed after $0.5$~ms of rotation of the south-pole C spinor $\xis$. Panel (a) schematically shows the orientations of the order parameter along the coordinate axes and the ring formed by the $\pi$ rotation of the initial state in the $xy$ plane. Preimages of south-pole and north-pole states are shown in panels (b) and (d), respectively. The color represents the rotation angle about the threefold symmetry axis $C_3$, which is pointing either (d)~up or (b)~down. Panels (c) and (e) show the isosurfaces $|\xi_1|^2 = 0.4$ and $|\xi_{-1}|^2 = 0.4$, respectively. All images correspond to $t=0.5$~ms after the start of the nonadiabatic creation ramp described in Sec.~\ref{sc:methods}.} 
\end{figure} 

The geometric representation of the C phase is shown in Fig.~\ref{fig:1}(b). The symmetry of the C phase corresponds to the largest discrete isotropy group among the spin-2 phases, namely, a tetrahedral group $T$. The correct cyclic order parameter space is therefore $G_{\textrm{C}} = \textrm{U}(1)\times\textrm{SO}(3)/T$. The C state possesses three twofold symmetry axes $C^{\prime\prime}_2$ and four threefold symmetry axes $C_3$, as shown in Fig.~\ref{fig:1}(b). Since one of the vertices of the tetrahedron points up, we will refer to such states as \emph{north-pole} states. Finally, we obtain the representative spinor of the so-called \emph{south-pole} spinor by a spin rotation through an angle of $\pi$ as
\begin{gather}
\xis =  
\exp \big{(}-i \pi F_x \big{)} \xin = (0, \sqrt{2/3}, 0, 0, \sqrt{1/3})^{\textrm{T}}.
\label{eq:C-s}
\end{gather}
The corresponding Majorana representation is shown in Fig.~\ref{fig:1}(f). We also note that the horizontal, vertical, north-pole, and south-pole states defined above refer to not only the representative spinors, but to any spinor obtained by continuous rotation of $\xih$, $\xiv$, $\xin$, or $\xis$ about the $z$ axis. 

\note{The third homotopy groups of the BN and the C order parameters are both known to be isomorphic to the group of integers, $\pi_3(G_{\textrm{BN}}) \cong \pi_3(G_{\textrm{C}}) \cong \mathbb{Z}$~\cite{KOBAYASHI2012577,Kaw2012.PhysRep520.253}. We therefore generalize the notion of skyrmions, originally defined as $\pi_3(S^3) \cong \mathbb{Z}$, to the appropriate topological space.}

\section{\label{sc:methods}Skyrmion creation in a trapped condensate}

To imprint the skyrmions, we apply the method used in Ref.~\cite{Hal2016.NatPhys12.478} for the experimental creation of a knot soliton in the polar phase of a spin-1 BEC~\cite{Kaw2008.PRL100.180403}. The $^{87}$Rb condensate is initially prepared in the spin state $\ket{m_\textrm{z} = 2}$ and then transferred to either the BN or the C phase using, for instance, two-photon Landau--Zener transitions~\cite{PhysRevA.72.053628}. 

The external magnetic field assumes the form $\mathbf{B}(\mathbf{r},t) = \mathbf{B}_{\textrm{q}}(\mathbf{r}) + \mathbf{B}_{\textrm{b}}(t)$, where $\mathbf{B}_{\textrm{q}}(\mathbf{r}) = B_{\mathrm{q}}^\prime (x\hat{\mathbf{x}} + y\hat{\mathbf{y}} - 2z\hat{\mathbf{z}})$ is a quadrupole magnetic field and $\mathbf{B}_{\textrm{b}}(t) = B_{\textrm{b}}(t)\hat{\mathbf{z}}$ is a uniform bias field which moves the field zero along the $z$ axis. We assume that initially $B_{\textrm{b}} \gg 2B_{\mathrm{q}}^\prime R_{\textrm{TF}}$, where $R_{\textrm{TF}}$ is the axial Thomas--Fermi radius of the condensate. This renders the spinor $\xi$ roughly uniform within the condensate prior to the skyrmion creation. In the first stage of the creation procedure, the bias field is ramped to zero in a highly nonadiabatic manner, ideally leaving the BEC unchanged~\cite{footnote_creation}. Subsequently, the spinor tends to rotate in the spin space such that
\begin{equation}\label{eq:spinrotation}
\begin{aligned}
\xi(\mathbf{r},t) = e^{-i\omega_{\textrm{L}}(\mathbf{r})t \ \hat{\mathbf{B}}_{\textrm{q}}(\mathbf{r})\cdot \mathbf{F}}\xi(\mathbf{r},0),
\end{aligned}
\end{equation}
where $\omega_{\textrm{L}}(\mathbf{r}) = \gF \muB |\mathbf{B}_{\textrm{q}}(\mathbf{r})|/\hbar$ is the position-dependent \mbox{Larmor} frequency, $\muB$ is the Bohr magneton, $\gF$ is the Land\'e $g$-factor, and $\hat{\mathbf{B}}_{\textrm{q}} = \mathbf{B}_{\textrm{q}}/|\mathbf{B}_{\textrm{q}}|$. The duration of the \mbox{Larmor} precession before imaging the BEC, $T_{\textrm{L}}$, is chosen such that the spinor assumes its initial orientation on the boundary of the condensate, where $\omega_{\textrm{L}} T_{\textrm{L}} = 2\pi$, and the enclosed volume can therefore be compactified into the 3-sphere $S^3$~\cite{Nak2003_book}. Inside this volume, the order parameter space is covered an integer number of times, which we identify as the mapping degree of a three-dimensional skyrmion. The Larmor precession corresponds to the rotations of the Majorana geometric configurations about the direction of the local magnetic field, ideally resulting in the structures shown schematically in Figs.~\ref{fig:2}(a) and ~\ref{fig:3}(a).

Equation~\eqref{eq:spinrotation} takes into account only the linear Zeeman coupling to the external magnetic field. In order to verify the feasibility of the proposed method in the presence of kinetic energy and interactions between the condensate atoms, we carry out three-dimensional simulations based on 
the spin-2 Gross--Pitaevskii equation
\begin{gather}\label{eq:GP}
i\hbar \frac{\partial}{\partial t}{\Psi}(\ve{r},t) = \mathcal{H}[\Psi]\Psi(\ve{r},t).
\end{gather}

The nonlinear Hamiltonian reads~\cite{Kaw2012.PhysRep520.253,footnote_inelastic}
\begin{equation}\label{eq:Hamiltonian}
\begin{aligned}
\mathcal{H}[\Psi] &= -\frac{\hbar^2 \nabla^2}{2m} + U(\mathbf{r}) + c_0n(\mathbf{r}) + \Big{[}\gF \muB \mathbf{B}(\mathbf{r},t) \\ &+ c_1n(\mathbf{r})\mathbf{S}(\mathbf{r})\Big{]} \cdot \mathbf{F} + c_2A_{20}(\mathbf{r})\mathcal{T},
\end{aligned}
\end{equation}
where the time-reversal operator $\mathcal{T}$ is defined in Eq.~\eqref{eq:timerev} and the spin-independent trapping potential is assumed to be cylindrically symmetric and harmonic, $U(\mathbf{r}) = m[\omega_{\rho}^2(x^2+y^2) + \omega_z^2 z^2]/2$, with $\omega_{\rho}$ and $\omega_z$ being the radial and axial trapping frequencies, respectively. The interaction constants are defined as
\begin{gather}
c_0 = \frac{4\pi\hbar^2}{m}\frac{4a_2 + 3a_4}{7},\textrm{ } c_1 = \frac{4\pi\hbar^2}{m}\frac{a_2 - a_4}{7},\nonumber\\
c_2 = \frac{4\pi\hbar^2}{m}\frac{7a_0 - 10a_2 + 3a_4}{7}, \label{eq:interactions}
\end{gather}
where $a_f$ is the $s$-wave scattering length corresponding to the scattering channel with total two-atom hyperfine spin $f$. 

We set the simulation parameters identical to those used in the experimental creation of knot solitons~\cite{Hal2016.NatPhys12.478}: The particle number is $N = 2.1 \times 10^5$, the optical trapping frequencies are $\omega_{\rho} = 2\pi \times 124$~Hz and $\omega_z = 2\pi \times 164$~Hz. \note{During the nonadiabatic creation ramp, the axial bias field is decreased from $10$~mG to zero in 60~$\mu$s and the quadrupole field gradient is kept constant at $B_{\mathrm{q}}^\prime=4.3$~G/cm.} The $s$-wave scattering lengths for $^{87}$Rb are $a_0 = 87.4\times\aB$, $a_2 = 92.4\times\aB$, $a_4 = 100.4\times\aB$, where $\aB=5.292 \times 10^{-11}$~m is the Bohr radius~\cite{Kem2002.PRL88.093201}. 

\note{To calculate the initial state of the condensate, we first find the natural ferromagnetic ground state using the successive overrelaxation algorithm. The ferromagnetic spinor is then instantaneously transferred to either $\xih$ or $\xis$, which simulates the two-photon Landau--Zener transition typically used in experiments. The subsequent dynamics are explored by numerically integrating Eq.~\eqref{eq:GP} with the Crank--Nicolson algorithm~\cite{CrankNicolson} and a time step of $2\times 10 ^{-4}/\omega_{\rho}$.} The simulated region is a cube of volume $(24a_r)^3$, where $a_r = \sqrt{\hbar/(m \omega_{\rho})} = 1.02$~$\mu$m. We use 200 grid points per dimension in order to keep the grid spacing significantly smaller than the condensate healing length.

\section{\label{sc:results}Results}

We apply the creation protocol described above to the initial biaxial nematic $\xih$ and cyclic $\xis$ states given by Eqs.~\eqref{eq:BN-h} and \eqref{eq:C-s}, respectively. Figure~\ref{fig:2} illustrates the skyrmion that appears after $0.5$~ms of 
the \mbox{Larmor} precession applied to a spin-2 BEC initially in the horizontal BN state. Figures~\ref{fig:2}(b) and \ref{fig:2}(d) show the preimages of the horizontal and vertical BN states with the color representing the rotation angle of geometric state about the $z$ axis. 
One can deduce the mapping degree directly from Figs.~\ref{fig:2}(b) and \ref{fig:2}(d). The preimage of the vertical state in Fig.~\ref{fig:2}(d), for example, consists of four loops; the geometric state rotates about its symmetry axis $C_2$ four times as one traverses each of four loops, resulting in the total mapping degree $Q_{\textrm{BN}}=16$. 
According to Eqs.~\eqref{eq:BN-h} and \eqref{eq:BN-v}, the preimages of the horizontal and vertical states are filled with $\ket{m_{\textrm{z}}=\pm 2}$ and $\ket{m_{\textrm{z}}=\pm 1}$, respectively. Therefore, the preimages of $\xih$ and $\xiv$ can be resolved from their component densities in the $z$-quantised basis, as shown in Figs.~\ref{fig:2}(c) and \ref{fig:2}(e), respectively. This establishes a means for the experimental detection of the skyrmionic structure directly from the spin-resolved images of the BEC. 

When applied to a BEC initially in the south-pole C state, an identical creation scheme to that described above generates the skyrmion illustrated in Fig.~\ref{fig:3}. The preimages of the south-pole [Fig.~\ref{fig:3}(b)] and north-pole [Fig.~\ref{fig:3}(d)] states are mostly filled with $\ket{m_{\textrm{z}}=1}$ and $\ket{m_{\textrm{z}}=-1}$ components, respectively, in agreement with Eqs.~\eqref{eq:C-s} and~\eqref{eq:C-n}. As one traverses each of the four rings in Fig.~\ref{fig:3}(d), the north-pole state rotates 
about the threefold symmetry axis $C_3$ six times. Therefore, the mapping degree of the C-state skyrmion is $Q_{\textrm{C}}=24$. 

\begin{figure}[h!]	
\includegraphics[width=0.9\columnwidth]{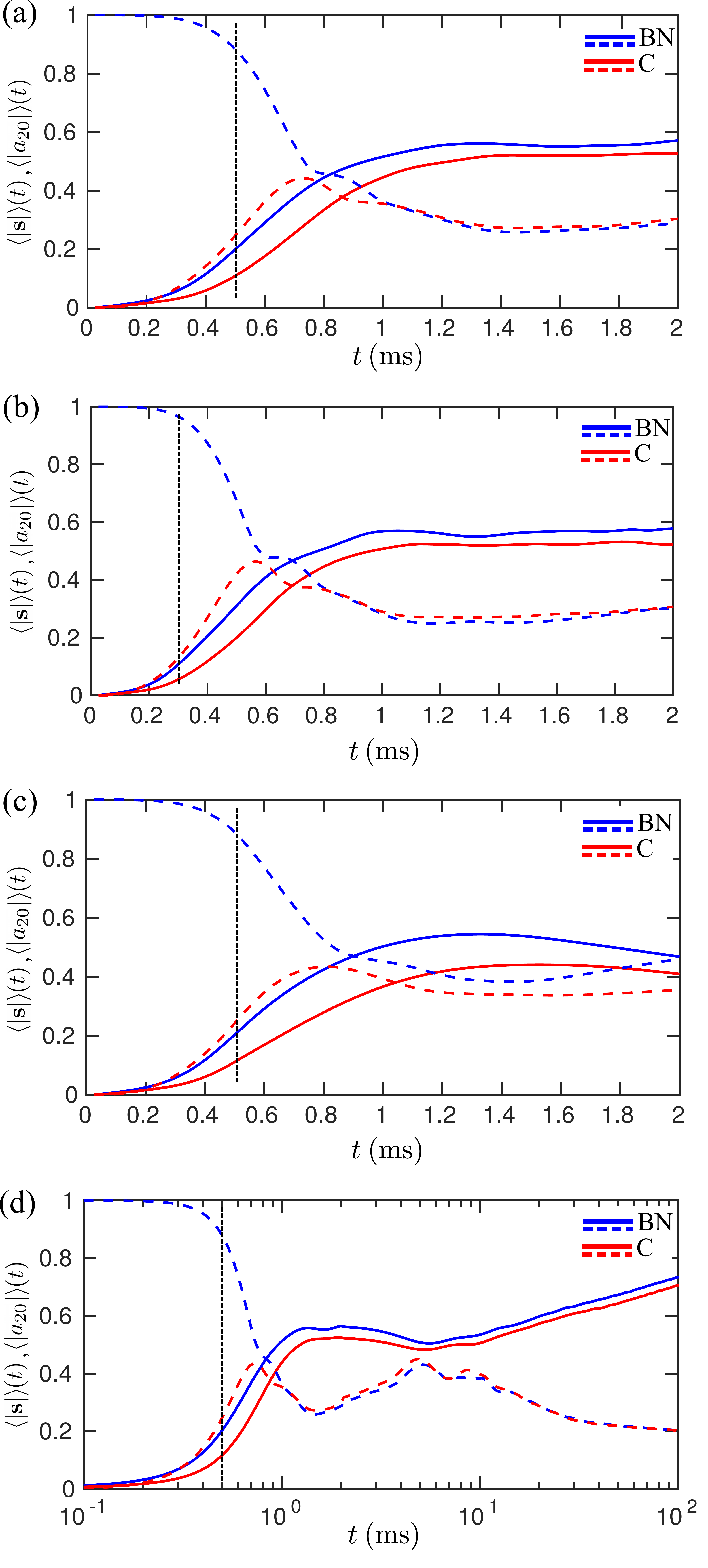}
\caption{\label{fig:4} (a) Expectation values of the average spin length $\ev{\mathbf{|s}(t)|}$ (solid lines) and the spin-singlet ampliture $\ev{|a_{20}(t)|}$ (dashed lines) for a condensate initially in either the BN (shown in blue) or the C (shown in red) state. Here $t=0$ corresponds to the beginning of the nonadiabatic creation ramp which places the field zero into the condensate and initiates the \mbox{Larmor} precession. The vertical line indicates the appearance of the skyrmions shown in Figs.~\ref{fig:2} and \ref{fig:3}. (b) As (a), but for stronger gradient, $B_{\textrm{q}}^{\prime}  = 8.6$~G/cm. \note{(c) As (a), but the quadrupole field is turned off after the creation ramp, and the skyrmions are left to evolve in a uniform magnetic field of $0.5$~G.} (d) As (a), but for a longer, logarithmic timescale.} 
\end{figure}

\note{Being topologically protected field configurations, the skyrmions are observed to be robust against small perturbations applied to the initial state. However, we emphasise that the skyrmion states are transitory states that appear and decay during the temporal evolution. We therefore study the stability of the underlying magnetic phases during the \mbox{Larmor} precession on timescales beyond the skyrmion creation time.}

Figure~\ref{fig:4}(a) illustrates the temporal evolution of the expectation values of the normalized spin magnitude,  $\ev{|\mathbf{s}|}(t) = \int n(\mathbf{r},t)|\mathbf{S}(\mathbf{r},t)| \intmeasure /(2N)$, and spin-singlet amplitude, $\ev{|a_{20}|}(t)  = \int n(\mathbf{r},t)|a_{20}(\mathbf{r},t)| \intmeasure/N $. We observe the destruction of the initial magnetic phase, which begins during the nonadiabatic creation ramp and continues gradually during the \mbox{Larmor} precession. For comparison, Fig.~\ref{fig:4}(b) shows the evolution of the system when the skyrmion is imprinted using a significantly stronger magnetic-field gradient than that depicted in Fig.~\ref{fig:4}(a). This causes the initial phase to be destroyed faster, but also enables a faster creation of the skyrmion due to the increased Larmor precession rate. As a result, the magnetic phase of the condensate is more accurately in the initial phase at the time the skyrmion is created when using a stronger gradient. 
\note{Qualitatively similar behaviour is observed when the quadrupole magnetic field is rapidly turned off after time $T_\mathrm{L}$ of Larmor precession and the skyrmion configuration is kept either in the identically zero magnetic field (data not shown) or in a uniform magnetic field of $0.5$~G that is ramped up during the switching off of the quadrupole field [Fig.~\ref{fig:4}(c)].}

In addition to the short-time dynamics, Fig.~\ref{fig:4}(d) shows the relaxation of the initial magnetic phases over timescales much longer than those required for skyrmion creation. Both the BN and C phases are observed to decay towards the ferromagnetic configuration when the quadrupole magnetic field is maintained. In contrast, the \note{initial uniform states $\xis$ and $\xih$} in the absence of the magnetic field gradient are observed to be stable during the whole time interval studied (data not shown), in good agreement with results of Ref.~\cite{Sch2004.PRL92.040402}.

\section{\label{sc:summary}Conclusions}

We have introduced exotic \note{skyrmion configurations} that can emerge \note{as transitory states} in the cyclic and biaxial nematic phases of three-dimensional spin-2 BECs. We simulated the creation of these three-dimensional skyrmions numerically for realistic parameter values corresponding to experimentally produced $^{87}$Rb condensates. We found that the created skyrmions are gradually destroyed during the temporal evolution of the trapped condensate due to the instability of the underlying initial magnetic phase. However, the lifetimes of the initial phases are long enough for the skyrmions to be imprinted and possibly detected in state-of-the-art experiments.

\begin{acknowledgments}
The authors gratefully acknowledge funding support from the Academy of Finland through its Centres of Excellence Program (Projects No.~251748 and No.~312300) and Grants No.~135794, No.~272806, and No.~308632, the Magnus Ehrnrooth Foundation, the Technology Industries of Finland Centennial Foundation, the National Science Foundation (Grant PHY-1519174), and Grants-in-Aid for Scientic Research (Grant No. 17K05554) from Japan Society for the Promotion of Science (JSPS). This project has received funding from the European Research Council (ERC) under the European Union's Horizon 2020 research and innovation programme under grant agreement No 681311 (QUESS). This work is also supported by JSPS and Academy of Finland  Research Cooperative Program (Grant No. 308071). We also  acknowledge CSC - IT Center for Science Ltd. (Project No. ay2090) and Aalto Science-IT project for provided computational resources.
\end{acknowledgments}

\bibliographystyle{apsrev4-1}
\bibliography{reflist,F2_Skyrmion_footnotes}

\begin{thebibliography}{58}
\bibitem [{foo({\natexlab{a}})}]{footnote_creation}%
  \BibitemOpen
  \bibinfo {note} {It may be experimentally preferable to turn the quadrupole
  field on during the creation ramp. Our additional numerical calculations show
  that both schemes result in identical skyrmionic structures.}\BibitemShut
  {Stop}%
\bibitem [{foo({\natexlab{b}})}]{footnote_inelastic}%
  \BibitemOpen
  \bibinfo {note} {Inelastic losses are not included in the Hamiltonian but are
  not expected to be important on the timescale of the skyrmion
  creation~\cite{Sch2004.PRL92.040402,PhysRevA.80.042704}.}\BibitemShut {Stop}%
\end{thebibliography}
\end{document}